# Initial results of finger imaging using Photoacoustic Computed Tomography

P. van Es[a], S. K. Biswas[a], H. J. Bernelot Moens[b], W. Steenbergen[a] and S. Manohar[a,*]

[a]University of Twente, Biomedical Photonic Imaging Group, MIRA Institute, P.O.Box 217, 7500 AE Enschede, The Netherlands.
[b]Ziekenhuis Groep Twente, Department of Rheumatology, Postbus 546, 7550 AM Hengelo, The Netherlands

We present a photoacoustic computed tomography investigation on a healthy human finger, to image blood vessels with a focus on vascularity across the interphalangeal joints. The cross-sectional images were acquired using an imager specifically developed for this purpose. The images show rich detail of the digital blood vessels with diameters between 100 μm and 1.5 mm in various orientations and at various depths. Different vascular layers in the skin including the subpapillary plexus could also be visualized. Acoustic reflections on the finger bone of photoacoustic signals from skin were visible in sequential slice images along the finger except at the location of the joint gaps. Not unexpectedly, the healthy synovial membrane at the joint gaps was not detected due to its small size and normal vascularization. Future research will concentrate on studying digits afflicted with rheumatoid arthritis to detect the inflamed synovium with its heightened vascularization, whose characteristics are potential markers for disease activity.

Keywords: Photoacoustic Tomography; rheumatoid arthritis; interphalangeal joints; finger vascularization.

**Address all correspondence to:** S.Manohar, University of Twente, TNW, Biomedical Photonic Imaging, Drienerlolaan 5, Enschede, The Netherlands, 7522DB; Tel: +31-(0)53-4893164; Fax: +31-(0)53-4891105; E-mail: s.manohar@utwente.nl

Rheumatoid arthritis (RA) is a chronic disease of the synovial joints affecting about 1% of the population.[1] The disease can be severely debilitating due to swelling, stiffness and pain associated with degradation of joints of the limbs and neck. The condition is characterized by inflammation, tissue proliferation and hypoxia in the synovium, which is the membrane between the joint capsule and joint cavity. Without treatment, these events lead to progressive cartilage and bone destruction.[1] The inflammation is maintained by the generation and relocation of blood vessels in the process of angiogenesis. [2] The cause of RA is not fully understood though it is known to be an autoimmune disease.[1]

Although there is no cure for RA, much progress has been made in its management using anti-inflammatory agents, synthetic or biological immunosuppressive drugs. [3] There is a need for sensitive imaging modalities that can facilitate reliable diagnosis at early stages and importantly, to monitor response to expensive therapeutic regimens.

It has been suggested in several studies that synovitis, manifested as thickening of the synovium, is a potential marker of disease activity and severity in RA. Imaging of the joints specifically to visualize synovitis is performed using ultrasound (US) imaging and Magnetic Resonance Imaging (MRI).[4, 5] US imaging allows visualization of synovium thickening and edema in progressed RA.[4] In early stages of disease, US imaging is less suitable due to low contrast between inflamed synovium and other soft tissue. Doppler US is used to assess the vascularity of the inflamed synovium but is observer dependent and lacks sensitivity to small blood vessels.[4] MRI provides good visualization of the inflamed synovium, but is expensive, largely inaccessible and requires contrast agents.[5]

Photoacoustic (PA) or optoacoustic imaging has the potential to address the shortcomings of these imaging methods.[6] PA combines high resolution ultrasound detection with the high optical absorption manifested by various chromophores in tissue.[7, 8] The method has been shown to be capable of imaging blood vessels based on optical absorption by hemoglobin (Hb) at various levels of oxygenation in blood. [7, 8] The group of X. Wang and Carson was first to apply the PA principle in the context of arthritis, with photoacoustic computed tomography (PACT) studies in small animals [9], on human cadaver fingers [10], and recently using PA in a linear geometry on fingers of healthy volunteers [11]. In the latter work, the authors, using a dual modality PA-US system, showed structural features in a joint identified with two different contrasts, while being able to delineate tendons from other soft tissue. Sun *et al* [12] performed PACT measurements on finger joints of subjects with osteoarthritis (OA), where PA intensities over the whole joint indicated correlation with inflammation, but individual blood vessels were not visualized. Large finger blood vessels were recently visualized in 3D by Ermilov *et al*. [13], largely to demonstrate technical feasibility.

In this article, we investigate PACT of a healthy human finger, to visualize the digital blood vessels focusing on vascularity across both interphalangeal joints. Such a study is required to help understand the requirements for ultimately visualizing the inflamed synovium in RA. A PACT setup was developed specifically to acquire *in vivo* cross-sectional images of the finger. The imager uses a 32-element curvilinear ultrasound array (Imasonic, Besançon, France) for detection of PA signals from the finger in backward mode, following its illumination with pulsed light from 6 optical fiber bundles. (Fig. 1) Each bundle has a diameter of 4 mm with an NA of 0.22. The detector array and optical fibers are fixed to the imaging tank which holds water. The tank and its contents rotate around the finger, which is immobilized between two Teflon rings at the center-of-rotation, thus acquiring multiple views around the object. Multiple slices along the finger can be acquired by stepping the imager through various heights while the finger remains stationary in the water. An acoustic filtered backprojection algorithm is used to reconstruct

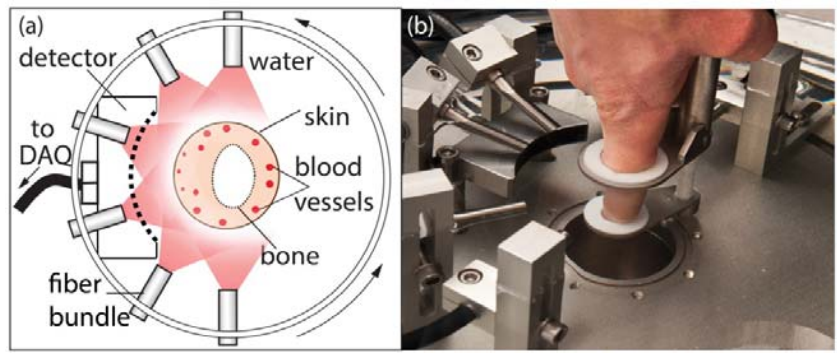

**FIG. 1** (a) Schematic overview, and (b) photograph of the photoacoustic computed tomography (PA-CT) setup. The finger is maintained stationary in water at the center of rotation of the curvilinear array, while illumination is provided by optical fibers in a backward-mode configuration.

the tomograms off-line. [14]

The ultrasound detector has a center frequency of 6.25 MHz and a -6 dB bandwidth of over 80%. [14, 15] with a measured minimum detectable pressure of 18 Pa. The array has a radius of curvature of 40 mm and covers 85 deg of a circle. The -3 dB elevation focus is roughly 1 mm at a distance of 48 mm, and the in-plane resolution for 12 views around the object is approximately 100 µm. A 32-channel pulser/receiver (Lecoeur-Electronique, Chuelles, France) sampling at 80 MSs$^{-1}$ is used for data acquisition. Illumination uses an Nd:YAG laser (Quanta-Ray pro 250, Spectra Physics, Mountain View, California) pumping an optical parametric oscillator (OPO, VersaScan- L532, GWU, Erftstadt, Germany) at 10 Hz. The radial and angular positions of the fiber bundles were adjusted to achieve roughly equal sized contiguous spots of 11 mm diameter at the skin surface. The pulse energy per fiber bundle was 6.5 mJ, giving an average fluence of 6.8 mJcm$^{-2}$ at the finger, well below the maximum permissible exposure (MPE) stated in IEC 60825- 1. Ultrasound images were made with a commercial ultrasound machine (Siemens Acuson S2000TM, Siemens AG, Erlangen, Germany).

Cross-sectional PA data from the healthy index finger were acquired from the subject (author S. Manohar) at 805 nm, by collecting 12 views with an angular step size of 30 deg. The total imaging time was ~ 1 minute per slice with 20 signal averages per view. Ten slices across the proximal interphalangeal (PIP) joint, and 10 slices across the distal interphalangeal (DIP) joint were acquired with a step size of 0.5 mm per slice.

A selection of 8 cross-sectional PA slices are shown in Fig. 2(a)-(h). The slices in Figs. 2(a)–2(c) are taken across the PIP joint, those in Figs. 2(d)–2(g) across the DIP Joint, and Fig.2(h) across the nail wall, at the approximate locations labeled in the US longitudinal section image [Fig. 2(j)]. A typical axial cross-section in US imaging is shown in Fig. 2(i). Figures 2(a#), 2(c#) etc. are magnifications of the boxed regions in the corresponding Figs. 2(a), 2(c) etc.

At 805 nm, hemoglobin and melanin are the dominant absorbers in the finger. In general, in all PA slices the skin surface layers and digital blood vessels are clearly visible. Blood vessels with diameters up to 1.5 mm and down to 100 to150 $\mu$m appear in the slices as circular cross-sections and can be traced from the base of the finger to the distal end. Some smaller blood vessels are observed as long thread like structures that primarily run parallel to the skin surface, and are best appreciated in the magnified images [Figs. 2(c#) and 2(g#)].

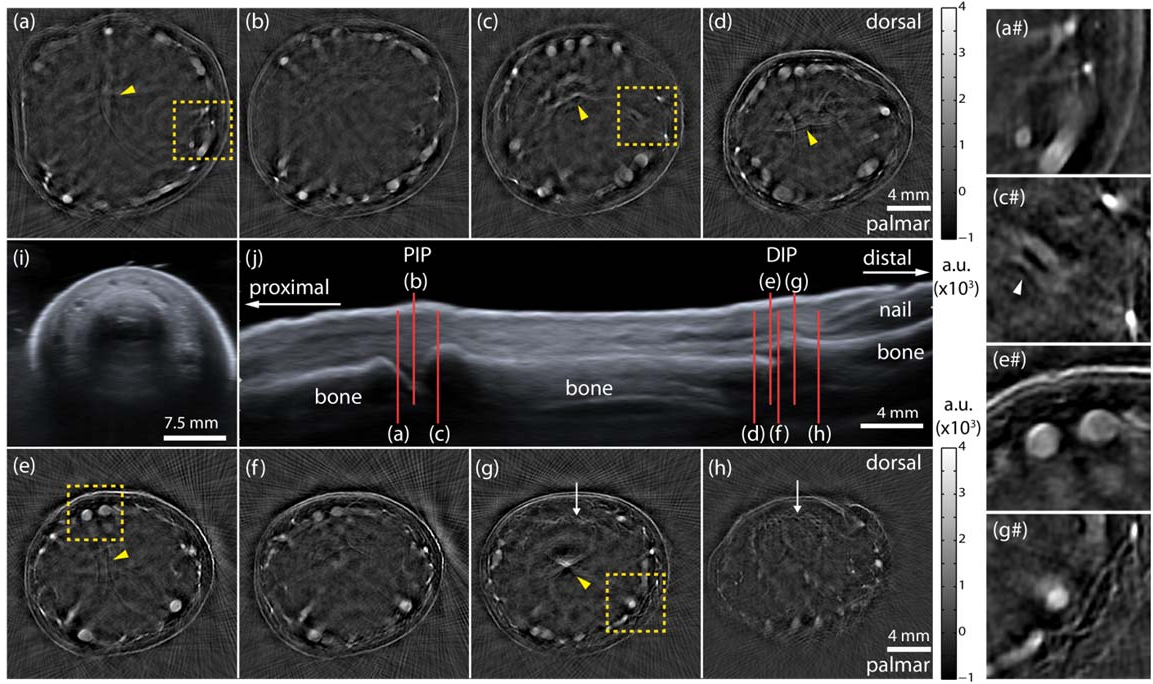

**FIG. 2** (a)-(h) Photoacoustic cross-sectional images along the healthy index finger of a volunteer showing blood vessels along the length of the finger as intense white regions. The images are taken at the positions shown in the longitudinal ultrasound image (j) and are concentrated at the proximal interphalengeal (PIP) joint and at the distal interphalengeal (DIP) joint. Bone appears white and blood vessels dark in (j) and (i) the cross-sectional ultrasound image close to the DIP. (a#) Enlarged image from box in (a) showing several blood vessels of various sizes. (c#) is an enlarged image of slice (c) with a blood vessel running towards the joint gap. (e#) shows two large blood vessels that run all the way to the nail bed. (g#) shows an enlargement of (g) showing blood vessels in the subdermis and dermis.

The majority of the large blood vessels are known to run through the hypodermis (Fig. 2(e#)) from which they branch to smaller blood vessels in the dermis [Figs. 2(c#) and 2(g#)]. It can also be observed that the density of fine blood vessels in the dermis on the palmar side increases towards the fingertip (compare Figs. 2(a)-(c) with Figs. 2(d)-(h) at 5 o' clock, and see detail in Fig. 2(g#)). These vessels are part of the dense sensory and thermoregulatory system in the finger tips. Another interesting observation is a fine and complex vascular pattern (arrows at 12 o' clock in Figs. 2(g) and 2(h)) distal to the DIP joint. This is most likely the superficial arcade of the nail wall that serves the nail bed vascularization further along the finger.

The skin layers visible in all slices are the epidermis and possibly the region below the epidermal-dermal junction, the former visible largely due to melanin and the latter due to the high density of capillaries of the subpapillary plexus. From the magnified Fig. 2(e#), the distance between the layers is approximately 300 $\mu$m.

The PA slices show artifacts especially at large depths due to reflections of PA skin and blood vessel signals on underlying bone (see arrow heads). No additional methods were applied to reduce the artifacts since they do not pose any problems in interpretation of the images. In fact, the relative presence/absence of these artifacts across sequential PA slice images can be used to advantage in order to ascertain the location of joint spaces. It is seen that these artifacts are absent or low in the slices [Figs. 2(b) and (f)] which are acquired approximately at the PIP and DIP joint gaps, respectively.

At the location of the joint gap the synovial membrane itself was not detected. This is not unexpected for healthy (non RA) finger joints since the vascularized membrane is less than approximately 50 $\mu$m thick with a normal density of blood vessels.[16] In the case of subjects with arthritic disease, inflammation will cause the thickening of the membrane due to proliferation of synovial tissue, and generation of blood vessels in angiogenesis. Although capillaries at the joint are clinically highly significant as these exchange gases and nutrients with the synovial fluid, their small sizes (<10µm) make them difficult to visualize. We expect that identification and monitoring of the capillary-rich synovial membrane as a whole are feasible and that changes in this membrane will be an early and sensitive indicator for inflammatory diseases.

To our knowledge these are the first *in vivo* PACT images showing the vascularization of the human index finger with such richness of detail. These results indicate that PA cross-sectional images might be used to detect vascular abnormalities such as those arising from angiogenesis associated with inflammatory rheumatic disease. Considering the resolution of the imager, we expect that the inflamed synovial membrane can be visualized when the thicknesses are greater than 100 $\mu$m. Whether the angiogenesis-driven optical absorption of the synovium in early disease will allow detection, requires systematic and comprehensive studies on subjects with RA.

This research is financially supported by the Netherlands Organization for Health Research and Development (ZonMw) under the program New Medical Devices for Affordable Health; and the High-Tech Health Farm Initiative of the Overijssel Center for Research and Innovation (OCRI). WS and SM are minority shareholders in PA Imaging BV, however the company did not financially support this research.

\*\*\*\*\*\*\*\*\*\*\*\*\*\*

**Caption List**

**Fig. 1** (a) Schematic overview, and (b) photograph of the photoacoustic computed tomography (PACT) setup. The finger is maintained stationary in water at the center of rotation of the curvilinear array, while illumination is provided by optical fiber bundles in a backward-mode configuration.

**Fig. 2** (a)-(h) Photoacoustic cross-sectional images along the healthy index finger of a volunteer showing blood vessels along the length of the finger as intense white regions. The images are taken at the positions shown in the longitudinal ultrasound image (j) and are concentrated at the proximal interphalengeal (PIP) joint and at the distal interphalengeal (DIP) joint. Bone appears white and blood vessels dark in (j) and (i) the cross-sectional ultrasound image close to the DIP. (a#) Enlarged image from box in (a) showing several blood vessels of various sizes. (c#) is an enlarged image of slice (c) with a blood vessel running towards the joint gap. (e#) shows two large blood vessels that run all the way to the nail bed. (g#) shows an enlargement of (g) showing blood vessels in the subdermis and dermis.